\documentclass[preprint, 3p, sort&compress]{elsarticle}

\usepackage{amsmath}
\usepackage{amssymb}
\usepackage{color}
\usepackage{todonotes}
\usepackage{pgfplotstable}
\usepackage[lofdepth,lotdepth]{subfig} % subfigure is deprecated; the options tell LaTeX
\usetikzlibrary{pgfplots.groupplots}

\journal{Computer Physics Communications}

\newcommand{\bout}{\texttt{BOUT++ }}

\newcommand{\bb}[1]{\mathbb{#1}}
\newcommand{\pd}[2]{\ensuremath{\frac{\partial #1}{\partial #2}}}
\newcommand{\pdd}[2]{\ensuremath{\frac{\partial^2 #1}{\partial #2^2}}}

\newcommand{\boutmatrix}[4]{\ensuremath{
    \left[
      \begin{array}{cc}
        #1 & #2 \\
        #3 & #4
      \end{array}
    \right]
}}

\renewcommand{\vec}[1]{\ensuremath{\mathbf{#1}}}

\newcommand*{\mysim}{\mathord{\sim}}

\pgfplotstableset{
  col sep=comma,
  create on use/ksp/.style={
    create col/expr={\pgfmathaccuma + \thisrow{KSP Iterations}},
  },
}

\definecolor{RYB1}{RGB}{128, 177, 211}
\definecolor{RYB2}{RGB}{251, 128, 114}
\definecolor{RYB3}{RGB}{253, 180, 98}
\definecolor{RYB4}{RGB}{179, 222, 105}
\definecolor{RYB5}{RGB}{141, 211, 199}
\definecolor{RYB6}{RGB}{255, 255, 179}
\definecolor{RYB7}{RGB}{190, 186, 218}

\definecolor{HTML1}{HTML}{2B83BA}
\definecolor{HTML2}{HTML}{D7191C}
\definecolor{HTML3}{HTML}{FDAE61}
\definecolor{HTML4}{HTML}{ABDDA4}

\pgfplotscreateplotcyclelist{bwcolors}{
  {black!20,very thick,mark size=1pt,mark=triangle*},
  {black!60,very thick,mark size=1pt,mark=square*},
  {black,very thick,mark size=1pt,mark=*},
}

\pgfplotscreateplotcyclelist{nicecolors}{
  {RYB1,very thick,mark size=1pt,mark=*},
  {RYB2,very thick,mark size=1pt,mark=square*},
  {RYB3,very thick,mark size=1pt,mark=triangle*},
}

\pgfplotscreateplotcyclelist{nicecolors-bar}{
  {RYB1!50!black,fill=RYB1},
  {RYB2!50!black,fill=RYB2},
  {RYB3!50!black,fill=RYB3},
}

\pgfplotscreateplotcyclelist{bwcolors-bar}{
  {fill=black},
  {fill=black!70},
  {fill=black!40},
  {fill=black!10},
}

\pgfplotscreateplotcyclelist{htmlcolors-bar}{
  {HTML1!50!black,fill=HTML1},
  {HTML2!50!black,fill=HTML2},
  {HTML3!50!black,fill=HTML3},
  {HTML4!50!black,fill=HTML4},
}

\pgfplotsset{
  every semilogy axis/.append style={
    legend style={
      draw=none,
      legend columns=-1,        % infinite columns = one row
      align=left,               % warning: this option enables multiple lines in the legend
      nodes={inner xsep=2ex},
    },
    table/x={Wallclock Time},
    table/y={Function Norm},
    enlarge x limits=0.15,      % this is for "buffer" space so as to not clip the "meta" data,
    table/meta=ksp,
    nodes near coords,          %=\rotatebox{90}{\pgfmathprintnumber\pgfplotspointmeta},
    every node near coord/.style={
      check for zero/.code={    % if meta=0, make the node a coordinate (which doesn't have text)
        \pgfmathfloatifflags{\pgfplotspointmeta}{0}{
          \pgfkeys{/tikz/coordinate}
        }{}
      },
      check for zero,           % call the 'check for zero' code
      above right,
      check for five digits/.code={ % a hack to move the numbers > 10,000 to the left
        \pgfkeys{/pgf/fpu=true}
        \pgfmathparse{\pgfplotspointmeta-10000}
        \pgfmathfloatifflags{\pgfmathresult}{+}{
          \pgfkeys{/tikz/left}
        }{}
        \pgfkeys{/pgf/fpu=false}
      },
      check for five digits,    % call the 'check for five digits' code
      check for ten/.code={     % a hack to position the '10' from no preconditioner plot
        \pgfkeys{/pgf/fpu=true}
        \pgfmathparse{\pgfplotspointmeta-10}
        \pgfmathfloatifflags{\pgfmathresult}{0}{
          \pgfkeys{/tikz/left}
        }{}
        \pgfkeys{/pgf/fpu=false}
      },
      check for ten,
    },
    point meta=explicit,
  },
  set scaling column/.code={
    \edef\xcol{\pgfkeysvalueof{/pgfplots/table/x}}
    \pgfplotstablegetelem{0}{\xcol}\of{\scalability}
    \edef\proc{\pgfplotsretval}
    \pgfplotstablegetelem{0}{\pgfkeysvalueof{/pgfplots/table/y}}\of{\scalability}
    \edef\offset{\pgfplotsretval}
    \pgfplotsset{table/y expr=\offset * \proc / \thisrowno{0}}
  },
  scaling/.style={
    forget plot, % don't add this plot to the legend; also makes it use the correct index
    opacity=0.4,
    dashed,
    set scaling column,
  },
}

\newcommand{\addscalingplot}[2][]{
  \addplot +[#1,scaling] table {#2};
  \addplot +[#1] table {#2};
}

\begin{document}

\begin{frontmatter}



\title{Improved Nonlinear Solvers in \bout}


\author[york]{Ben Dudson}
\author[iit]{Sean Farley}
\author[mcs]{Lois Curfman McInnes}
\address[york]{York Plasma Institute, Department of Physics\\
University of York\\
York YO10 5DD, United Kingdom\\
\mbox{}\\
}
\address[iit]{Mathematics Department\\
Illinois Institute of Technology\\
Chicago, IL 60616, USA\\
\mbox{}\\
}
\address[mcs]{Mathematics and Computer Science Division\\ 
Argonne National Laboratory\\
Argonne, IL 60439, USA
}

\begin{abstract}
  Challenging aspects of large-scale turbulent edge simulations in plasma physics include
  robust nonlinear solvers and efficient preconditioners.  This paper presents recent
  advances in the scalable solution of nonlinear partial differential equations in BOUT++,
  with emphasis on simulations of edge localized modes in tokamaks. A six-field,
  nonlinear, reduced magnetohydrodynamics model containing the fast shear Alfv\'en wave
  and electron and ion heat conduction along magnetic fields is solved by using Jacobian-free
  Newton-Krylov (JFNK) methods and nonlinear GMRES (NGMRES). Physics-based preconditioning
  based on analytic Schur factorization of a simplified Jacobian is found to result in an
  order of magnitude reduction in runtime over unpreconditioned JFNK, and NGMRES is shown
  to significantly reduce runtime while requiring only the nonlinear function
  evaluation. We describe in detail the preconditioning algorithm, and we discuss parallel
  performance of NGMRES and Newton-Krylov methods using the PETSc library.
\end{abstract}

\begin{keyword}
physics-based preconditioning, edge localized modes, Newton-Krylov, nonlinear GMRES
\end{keyword}

\end{frontmatter}

\section{Introduction}
\label{sec:introduction}

The international community is committed to the development of fusion energy in order to
meet growing world energy needs.  This is a complex, international, and costly endeavor;
the next experimental magnetic-fusion plasma confinement device,
ITER~\cite{iter:homepage}, is estimated to cost in excess of \$10B.  The handling of the
plasma exhaust is a challenge for the operation of ITER and for the design of the future
demonstration power plant DEMO. To achieve its design goals, ITER must operate in a
high-performance regime with steep pressure gradients close to the plasma edge
\cite{keilhacker1984,doyle2008}. While this capability improves the plasma performance,
the steep pressure gradients and associated bootstrap current can destabilize
peeling-ballooning modes \cite{connor-1998,snyder-2002,wilson-2002}. These result in
eruptions from the plasma edge, known as edge localized modes (ELMs) (see, e.g.,
\cite{maggi2010}). Scaling from existing experiments indicates that these violent events,
if uncontrolled, could produce peak power loads of around 1 GW/m$^2$ onto material
surfaces in ITER \cite{leonard1999,federici2003} and potentially more in a DEMO power
plant.  Such loads would result in unacceptable erosion of plasma facing components. Thus,
researchers are extremely interested in understanding and controlling ELMs, and work is
ongoing to use several nonlinear simulation codes in order to simulate ELMs, predict their
size in ITER, and find ways to mitigate them.

Nonlinear simulations of tokamak plasmas are challenging because of the wide range of
length scales and timescales present: The characteristic ion cyclotron motion has scales
around $1$ mm and $10^{-7}$ seconds, while the machine has length scales of $\mysim 10$ m
and transport timescales in seconds. If electron physics is important, then the
characteristic scales become smaller by the square root of the mass ratio, a factor of
$\mysim 60$.  These small cyclotron scales are due to the strong magnetic field (typically $1
- 5$ Teslas), which is used to confine the plasma in a tokamak. This produces strong
anisotropy in the system, as particles are confined perpendicular to the magnetic field
but are essentially free to move along it, resulting in a difference of $10^8$ in heat
conduction rates perpendicular and parallel to the field. The topology of the magnetic
field poses challenges for simulation, as the core fieldlines form helices with a pitch
angle that varies across the device. At the boundary of the core plasma, in the region we
are interested in simulating, a separatrix is formed, outside of which the fieldlines
intersect material boundaries rather than forming closed toroidal ``flux'' surfaces.
Furthermore, the edge region of tokamaks is complicated by the interaction with neutral
gas and material surfaces, bringing in atomic physics, sputtering, and deposition
processes.

Analytic reduction of the full problem, employing asymptotic expansions in small
parameters such as the ratio of the Larmor radius to machine size, has been successfully
used for many years (see, e.g., \cite{krommes2012} and references therein). In this paper
we consider fluid equations, which evolve moments of the particle velocity distribution,
and remove the smallest scales associated with cyclotron motion \cite{braginskii1965}.
Further simplifications include reduced magnetohydrodynamics (MHD), exploiting the
anisotropy of the system in which length scales parallel to the magnetic field are much
larger than those perpendicular to the field (see, e.g., \cite{strauss1976,
  hazeltinemeiss}).  By assuming force balance, this approach allows us to analytically
remove the fast magnetosonic wave, which would be the fastest wave in our system. This
wave is not important for the relatively slow MHD instabilities we wish to study here, but
it is responsible for establishing force balance along the magnetic field. After this
reduction, the system of equations still contains two fast timescales that cannot be
removed without also removing physics of interest: the shear Alfv\'en wave and the
parallel heat transport. These can severely limit the timestep in numerical simulations,
so implicit and semi-implicit methods are widely used in plasma simulation
\cite{jardin2012}.  In BOUT++ edge simulations of ELMs \cite{dudson2009bout,xu2010prl,
  dudson2011ppcf,xu2011nf}, such techniques help address challenges in the broad range of
temporal scales, complicated geometry with a range of spatial scales, and competing
physical effects of the various degrees of freedom.

This paper focuses on efficient and scalable nonlinear solvers for BOUT++ ELM simulations.
We construct a \emph{physics-based} preconditioner that reduces the time to solve
nonlinear systems with Jacobian-free Newton-Krylov (JFNK) techniques~\cite{knollkeyes:04}.  The
term \emph{physics-based preconditioner} has been used to describe various approaches to
application-specific preconditioners that incorporate insight into the physics behind the
equations being solved \cite{MousseauKnoll03,ChaconKnollEtAl02,SantilliScotti11,
  PremnathPattisonEtAl09,HirshmanSanchezEtAl11,jamroz2010,McCourtRognlienEtAl12}.
Inspired by the approach of Chac\'on~\cite{chacon2008}, we use the term here to describe
our development of a custom preconditioning algorithm based on analytic reduction and Schur
factorization of the system Jacobian. Identifying the key physics terms allows a
simplification of the resulting equations and enables efficient numerical solution.  A
similar approach has been previously attempted within the NIMROD code \cite{jamroz2010},
where a Crank-Nicholson (CN) Jacobian-free Newton-Krylov scheme was used along with
physics-based preconditioning. In that work numerical instability was encountered and
attributed to the CN scheme. In this paper we exploit the field-aligned coordinate system
used in BOUT++~\cite{dudson2009bout} to construct a preconditioner that reduces a large 3D
problem to the solution of many one-dimensional systems, each of which can be solved
independently. This approach greatly improves convergence of Newton-Krylov methods in
implicit timestepping schemes, without causing numerical instability.

We also introduce the complementary approach of nonlinear GMRES \cite{Oosterlee:2000tg},
which offers the advantage of requiring only nonlinear function evaluations from the
application, rather than also needing Jacobian-vector products and preconditioning for
Newton-based methods.  We demonstrate that both of these approaches---physics-based
preconditioned Newton-Krylov and nonlinear GMRES---significantly reduce the time for
solving the nonlinear ELM systems at each timestep compared with a Newton-Krylov approach
with no preconditioning, and we demonstrate good strong scaling on up to 512 processor
cores.

The remainder of this paper explains the details of our approach.  Section
\ref{sec:BOUT++} introduces the BOUT++ application and its approach to modeling ELMs.
Section \ref{sec:nonlinear_solvers} describes the algorithms and software that we use to
solve the resulting nonlinear systems that arise at each timestep, including both
Newton-Krylov methods with physics-based preconditioning and nonlinear GMRES.  Section
\ref{sec:results} presents experimental results, while Section \ref{sec:conclusions}
discusses conclusions and directions for future work.

\section{BOUT++ Simulation Overview}
\label{sec:BOUT++}

The numerical methods studied in this paper are quite general, but are applied here to the
solution of reduced MHD for fusion applications within the BOUT++ framework.

\subsection{\bout Capabilities}
\label{subsec:boutsimcode}

BOUT++, a parallel MPI + OpenMP finite
difference code written in C++ \cite{dudson2009bout}, simulates a range of plasma
fluid equations in curvilinear coordinate systems---in particular, reduced MHD models in
coordinates aligned with a strong background magnetic field. Current applications of
BOUT++ include the study of ELMs \cite{xu2010prl,dudson2011ppcf,xu2011nf}, plasma
turbulence \cite{friedman2012arxiv}, and blob dynamics \cite{angus2012prl}. In this paper
BOUT++ is extended to more efficiently simulate heat conduction parallel to magnetic
fields, an important process in all these areas. Parallel heat conduction modifies the
linear structure and growth rate of plasma instabilities and plays a crucial role in the
loss of energy to material surfaces in tokamak edge simulations of turbulence and ELMs.

\subsection{Reduced MHD Equations}
\label{subsec:equations}

Since the linear stability threshold is well described by ideal MHD~\cite{snyder-2002}, it
is reasonable to start studying ELMs by using fluid models based on MHD. Several initial-value codes
including BOUT++~\cite{dudson2011ppcf}, M3D-C$^1$~\cite{ferraro2010} and NIMROD~\cite{burke2010}, 
have now been benchmarked and shown good agreement with linear ideal MHD codes ELITE~\cite{snyder-2002,wilson-2002} and
GATO~\cite{bernard1981}. 
Nonlinearly, ideal MHD theory predicts
narrowing structures erupting outwards, leading to a finite time singularity
\cite{wilson-2004}.  Long before this happens, additional effects must become important in
order to allow plasma to decouple from magnetic fields and relax to a new
equilibrium. Nonlinear modeling of ELMs has therefore concentrated on fluid models
incorporating a range of different nonideal effects \cite{burke2010,pamela2010,
  hoelzl2012,sugiyama2009,jardin2007,ferraro2010}.

Since each simulation code currently uses different fluid models, nonlinear benchmarking
of codes is difficult. In order to enable future comparisons between codes, a set of
equations close to those used in JOREK~\cite{huysmans2009} has been implemented in BOUT++,
which we will refer to in this paper as the six-field model.  Six scalar fields are
evolved: mass density $\rho$; electron and ion temperatures $T_e$ and $T_i$, respectively;
vorticity $U=\vec{b}_0\cdot\nabla\times\vec{v} \simeq \frac{1}{B_0}\nabla_\perp^2\phi$;
parallel velocity $v_{||} = \vec{b}_0\cdot\vec{v}$; and parallel vector potential $A_{||} =
\vec{b}_0\cdot\vec{A}$. From these auxiliary quantities are calculated the total plasma
pressure $P = \rho (T_e + T_i)$ and parallel current density $J_{||} =
\vec{b}_0\cdot\vec{J} = -\nabla_\perp^2 A_{||}$.  The equilibrium magnetic field unit
vector is $\vec{b}_0=\vec{B}_0/B_0$. Here all quantities with subscript ``0'' are
calculated from the starting equilibrium and are not evolved. The evolving quantities are
deviations from equilibrium; for example, the total mass density is $\rho_0 + \rho$. The
equations for plasma density and temperatures are
\begin{align}
  \pd{\rho}{t} &= -\vec{v} \cdot \nabla \left(\rho + \rho_0\right)
                + \left(\nabla \cdot \vec{v}\right) \left(\rho + \rho_0\right)
                + D_\perp\nabla_\perp^2\rho \cr
  \pd{T_s}{t} &= -\vec{v} \cdot \nabla \left(T_s + T_{s0}\right)
               - \frac{2}{3} \left(\nabla \cdot \vec{v}\right) \left(T_s + T_{s0}\right) \cr
              &+ \frac{1}{\rho + \rho_0} \left[
                   \nabla_{||} \cdot \left(\chi_{s||}\partial_{||}T_s\right)
                   + \chi_{S\perp}\nabla_\perp^2T_s
                 \right] + \frac{2}{3\left(\rho+\rho_0\right)} W_s ,
  \label{eq:6field1}
\end{align}
where the total plasma velocity $\vec{v} = \vec{v}_{E\times B} + \vec{b} v_{||}$ is given
by an $E\times B$ drift perpendicular to magnetic fieldlines $\vec{v}_{E\times B} =
\frac{1}{B_0}\vec{b}_0\times\nabla\phi$ and parallel flow $v_{||}$ along them. The above
equations describe the advection ($\vec{v}\cdot \nabla$) and compression ($\nabla \cdot
\vec{v}$) of density and temperatures ($s = {e,i}$), collisional diffusion perpendicular
to the magnetic field with coefficients $D_\perp$ and $\chi_{s\perp}$, and parallel
conduction of heat along magnetic fields with coefficient $\chi_{s||}$. Derivatives are
split into those parallel to the magnetic field $\nabla_{||} = \vec{b}_0\cdot\nabla$ and
those perpendicular $\nabla_\perp = \nabla - \vec{b}_0\cdot\nabla$.

The velocity of the plasma is described by the vorticity and parallel flow:
\begin{align}
  \pd{U}{t} &= -\vec{v}\cdot\nabla U + \frac{1}{\rho+\rho_0} \left[B_0^2\nabla_{||}
                \left(\frac{J_{||} + J_{||0}}{B_0}\right) \right. \cr
            &+ \left. 2\vec{b}_0 \times \vec{k}_0 \cdot \nabla P
             + \nu_{||}\partial_{||}^2U + \nu_\perp\nabla_\perp^2U \right] \cr
  \pd{v_{||}}{t} &= -\vec{v}\cdot\nabla v_{||} - \frac{1}{\rho+\rho_0}\nabla_{||} P .
  \label{eq:6field2}
\end{align}
The vorticity equation (evolving $U$) contains a term responsible for shear Alfv\'en wave
propagation ($B_0^2\nabla_{||}\frac{J_{||}}{B_0}$), instability drive due to equilibrium
magnetic field curvature $\kappa$ and pressure gradient $\nabla P$, as well as parallel
and perpendicular viscosity terms $\nu_{||}$ and $\nu_\perp$.

The evolution of the magnetic field is described by Ohm's law for the electric field
parallel to the magnetic field $\vec{b}\cdot\vec{E} = \vec{b}\cdot\eta\vec{J}$,
\begin{equation}
  \pd{A_{||}}{t} = -\nabla_{||}\phi - \eta J_{||} ,
  \label{eq:6field3}
\end{equation}
where $\eta$ is the parallel resistivity.

\subsection{Timescales for Shear Alfv\'en Wave and Parallel Heat Conduction}
\label{subsec:alfvenwave}

The timescales of interest in equations~\eqref{eq:6field1}-\eqref{eq:6field3} are those of
the linear growth of peeling-ballooning modes and nonlinear eruption of filaments,
typically of the order of $\mysim 100\mu$s up to milliseconds. The above set of equations
contains timescales much faster than this, which must be preconditioned in an implicit
scheme.  The vorticity and $A_{||}$ equations contain the terms
\[
  \pd{U}{t} =  \frac{B_0^2}{\rho_0}\nabla_{||} \left(\frac{J_{||}}{B_0}\right)
  \qquad
  \pd{A_{||}}{t} = -\nabla_{||}\phi ,
\]
which reduce to
\[
  \pdd{\phi}{t} = \nabla_\perp^{-2} \left[
                    \frac{B_0^3}{\rho_0}\nabla_{||} \left(\nabla_\perp^2\nabla_{||}\phi / B_0\right)
                  \right]
           \simeq \frac{B_0^2}{\rho_0}\nabla_{||}^2 \phi ,
\]
corresponding to the shear Alfv\'en wave traveling along magnetic fieldlines. Given the
strong magnetic field $B_0$ and low mass density $\rho_0$ in tokamaks, wave speeds of $v_A
\mysim 10^7$m/s are typical, which for parallel grid spacing of $\mysim 10$ cm $ - 1$ m give
CFL timesteps $\mysim 10^{-8}$s, 3 to 4 orders of magnitude shorter than timescales of
interest.

The second fast timescale is introduced through the ion and electron temperature equation
\[
  \pd{T_s}{t} = \frac{1}{\rho_0}\nabla_{||}\left(\chi_{s||}\partial_{||}T_s\right) ,
\]
which is a diffusion equation along magnetic fieldlines. At the high temperatures relevant
to fusion, electron thermal speeds along fieldlines of $10^6-10^7$m/s are typical, and
hence parallel thermal diffusion is also fast compared with nonlinear evolution
timescales.

The above equations contain one other wave along magnetic fieldlines: the sound wave
described by
\[
  \pd{v_{||}}{t} = -\frac{1}{\rho_0}\nabla_{||} P \qquad
  \pd{\rho}{t} = -\rho_0\nabla_{||}v_{||} \qquad
  \pd{T_s}{t} = -\frac{2}{3}T_{s0}\nabla_{||}v_{||} .
\]
The speed of this wave is typically around $10^5$ m/s for the problems in which we are
interested, several orders of magnitude slower than the Alfv\'en wave and parallel
electron heat conduction. It is trivial to extend the preconditioner described below to
include this wave, but this extension was found to make little difference to the
convergence rate and so is not included here.

In the following sections we outline the implicit timestepping method and construct a
preconditioning scheme that can be applied to address both the Alfv\'en wave and
temperature conduction timescales.

\section{Scalable Nonlinear Solvers}
\label{sec:nonlinear_solvers}

The coupled set of reduced MHD equations~\eqref{eq:6field1}-\eqref{eq:6field3} are
discretized by the method of lines, using 4th-order central differencing for diffusive terms,
and 3rd order WENO \cite{jiang-1996} for upwinding. Because this work focuses on nonlinear
solvers, we chose the timestepping algorithm to be the same for each test case, namely, a
variable order backward differential formulation (BDF)~(see, e.g., \cite{Ascher_B1998}).
This family comprises methods that are implicit and linear multistep, and that allow
variable timestepping.  At each timestep, by writing the system state as a vector, we
obtain from equations~\eqref{eq:6field1}-\eqref{eq:6field3} a nonlinear system of the form
\begin{equation}
f(u) = f \left( \begin{array}{c}
  \rho \\
  T_i \\
  T_e \\
  U \\
  v_{||} \\
  A_{||}\end{array}\right) = 0 ,
\label{eq:f=0}
\end{equation}
where $f: \, \bb{R}^n \to \bb{R}^n$.
This requires the solution of a nonlinear system, here using Newton-Krylov and NGMRES
methods.

\subsection{Newton-Krylov Methods}
\label{subsec:newtonprecon}

In a Newton-Krylov method, we solve the nonlinear system~\eqref{eq:f=0} at each timestep
using a Newton iteration (see, e.g., \cite{NW99}),
\[
  u_{k+1} = u_k - [ f'(u_k)]^{-1} f(u_k), \;\; k=0,1, \ldots.
\]
Here $u_0$ is an initial approximation to the solution, and $f'(u_k)$, the Jacobian, is
nonsingular at each iteration. A Krylov iterative method (here GMRES) is then used to
solve this linear system at each Newton iteration.  In practice, the Newton iteration is
implemented by the following two steps:
\begin{enumerate}
  \item (Approximately) solve \;\; $f'(u_k) \Delta u_k = -f(u_k)$.
  \item Update \;\; $u_{k+1} = u_k + \alpha \Delta u_k$.
\end{enumerate}
Here $0 < \alpha \leq 1$ is a scalar. In these experiments, we use a JFNK variant, where
Jacobian-vector products are computed matrix-free~\cite{knollkeyes:04} by an approximation
\[
  f'(v) a \approx \frac{f(v + h \cdot a) - f(v)}{h},
\]
for a differencing parameter $h$.  By computing finite-difference Jacobian-vector products
for the iterative solve, assembling the actual Jacobian is never required.

If the nonlinear system being evolved contains a wide range of timescales, as in most
plasma simulations, then it is said to be ``stiff,'' and the Jacobian can be
ill-conditioned. In this case the Krylov method will require an unacceptably large number
of iterations.  Preconditioning the linearized Newton systems, which have the form
$\mathbb{J} x = b$, where $\mathbb{J}$ is the nonsingular $n \times n$ Jacobian matrix and
$x$ and $b$ are $n$-dimensional vectors, transforms these to equivalent systems
$\mathbb{P} \mathbb{J} x = \mathbb{P} b$ through the action of a preconditioner,
$\mathbb{P}$, whose inverse action approximates that of the Jacobian but at smaller cost.
Choosing an effective approximation is one of the goals of this work, since the
preconditioning phase is crucial to achieving low computational cost and scalable
parallelism.

\subsection{Physics-Based Preconditioning}
\label{subsec:preconditioning}

Physics-based preconditioning \cite{chacon2008,chacon02,ChaconKnollEtAl02} is a method by
which insight into the physics behind the equations being solved is used to precondition
the Newton-Krylov method.  After first identifying the stiff terms in the equations, an
approximate inverse is constructed with block Gaussian elimination (Schur factorization) of
the simplified analytic Jacobian.  The resulting matrices have a parabolic form, even if
the starting system is hyperbolic, as a result of the implicit timestepping
procedure. Multigrid methods are often used to solve these systems (see, e.g.,
\cite{chacon2008}), but these are difficult to implement in complex geometries and an
existing codebase that was not designed with them in mind. Here we describe a simpler
scheme that can be used to efficiently solve the factorized Jacobian in field-aligned
coordinates without the complexity of a multigrid method.

To derive a preconditioner, we begin by simplifying the nonlinear six-field system,
equations~\eqref{eq:6field1}-\eqref{eq:6field3}, retaining the linear vorticity terms
driving perpendicular plasma $E\times B$ motion, the fast parallel Alfv\'en wave that
propagates the motion along fieldlines, the linear terms describing perpendicular
advection of density and temperature, and the parallel diffusion of temperature
perturbations. This procedure reduces the equations to
\begin{align}
  \pd{\rho}{t}   &\simeq -\vec{v}_{E\times B} \cdot \nabla_\perp \rho_0 \cr
  \pd{T_s}{t}    &\simeq -\vec{v}_{E\times B} \cdot \nabla_\perp T_{s0}
                 + \frac{1}{\rho_0} \nabla_{||} \cdot \left(\chi_{s||}\partial_{||}T_s\right) \cr
  \pd{U}{t}      &\simeq \frac{1}{\rho_0} \left[B_0^2\nabla_{||} \left(\frac{J_{||}}{B_0}\right)
                 + 2\vec{b}_0 \times \vec{k}_0 \cdot \nabla P\right] \cr
  \pd{v_{||}}{t} &\simeq 0 \cr
  \pd{A_{||}}{t} &\simeq -\nabla_{||}\phi .
  \label{eq:simplified}
\end{align}

We have therefore neglected all nonlinear terms (those involving two or more evolving
quantities), all perpendicular diffusive terms, viscosity, and resistivity. The neglect of
nonlinear terms is justified because we wish to follow the nonlinear dynamics and thus
should not step over these timescales; instead, we focus on preconditioning the fast
linear dynamics. Perpendicular diffusion and resistive effects are also slow relative to
the parallel dynamics and so can be dropped.  We stress that these simplifications made in
the preconditioner do not affect the solution to the full six-field model, only the
convergence rate toward it.  When the assumptions made here are violated, the
preconditioner becomes less effective, but the solution is not compromised (provided that
the Krylov method still converges and does not stall).

In an implicit timestepping scheme (Section~\ref{subsec:newtonprecon}) the operator to be
inverted is $\mathbb{A} = \mathbb{I} - \gamma\mathbb{J}$, where $\mathbb{J}$ is the system
Jacobian; $\gamma$ is usually proportional to the timestep and is determined by the choice
of integration scheme. For preconditioning, a simplified form of the Jacobian is derived
from Equations~\eqref{eq:simplified} and is given in Equation \eqref{eq:jacobian}.

{\footnotesize
\begin{equation}
  \bb{J} = \left[
    \begin{array}{c|c|c|c|c|c}
      0 & 0 & 0 & 0 & 0 & \frac{1}{B}\vec{b}\times\nabla\rho_0\cdot\nabla\nabla_\perp^{-2}[B_0 \\

      0 & \frac{\chi_{i,||}}{\rho_0}\nabla_{||}^2 & 0 & 0 & 0 &
      \frac{1}{B}\vec{b}\times\nabla T_{i0}\cdot\nabla\nabla_\perp^{-2}[B_0 \\

      0 & 0 & \frac{\chi_{e,||}}{\rho_0}\nabla_{||}^2 & 0 & 0 &
      \frac{1}{B}\vec{b}\times\nabla T_{e0}\cdot\nabla\nabla_\perp^{-2}[B_0 \\

      0 & 0 & 0 & 0 & 0 & 0 \\

      0 & 0 & 0 & 0 & 0 & -\nabla_{||}\nabla_\perp^{-2}[B_0 \\

      \frac{2}{\rho_0}\vec{b}\times\kappa\cdot\nabla\left[\left(T_{e0}+T_{i0}\right)\right. &
      \frac{2}{\rho_0}\vec{b}\times\kappa\cdot\nabla\left[\rho_0\right. &
      \frac{2}{\rho_0}\vec{b}\times\kappa\cdot\nabla\left[\rho_0\right. &
      0 &
      -\frac{1}{\rho_0}B_0\nabla_{||}\nabla_\perp^2 &
      0
    \end{array}
  \right]
\label{eq:jacobian}
\end{equation}
}

Following the methodology of \cite{chacon2008}, we first factorize the matrix
$\bb{A}=\bb{I} - \gamma\bb{J}$ using Schur factorization:
\begin{equation}
  \bb{A}^{-1} = \left( \bb{I} - \gamma\bb{J} \right)^{-1} =
  \boutmatrix
    {\bb{E}} {\bb{U}}
    {\bb{L}} {\bb{D}}^{-1} =
  \boutmatrix
    {\bb{I}} {\bb{E}^{-1}\bb{U}}
    {0} {\bb{I}}
  \boutmatrix
    {\bb{E}^{-1}} {0}
    {0} {\bb{P}_{Schur}^{-1}}
  \boutmatrix
    {\bb{I}} {0}
    {-\bb{L}\bb{E}^{-1}} {\bb{I}},
\label{eq:factorisation}
\end{equation}
where $\bb{L}$ is a row vector containing the vorticity drive terms due to curvature and
parallel current, $\bb{U}$ is a column vector containing the linear $E \times B$ advection
terms, and $\bb{E}$ contains the parallel heat conduction terms, which have the form
$\bb{I} - \gamma \frac{\chi_{||e,i}}{\rho_0}\nabla^2_{||}$.

The matrix $\bb{P}_{Schur} = \bb{D} - \bb{L}\bb{E}^{-1}\bb{U}$ is the Schur complement;
and since nonlinear terms are neglected, $\bb{D}$ is the identity matrix.  To simplify the
form of $\bb{P}_{Schur}$, we drop terms involving only perpendicular derivatives, leaving
only the shear Alfv\'en wave:
\[
  \bb{P}_{Schur} \simeq \bb{I} - \gamma^2\frac{B_0^2}{\rho_0}
  \nabla_{||}\nabla_\perp^2\nabla_{||}\nabla_\perp^{-2} .
\]
In order to remove the perpendicular derivatives, the following approximation is used:
\[
  \nabla_{||}\nabla_\perp^2\nabla_{||}\nabla_\perp^{-2}
  \simeq
  \nabla_{||}^2 -\frac{2\nabla_{||}\left(RB_\theta\right)}{RB_\theta} ,
\]
where $R$ and $B_\theta$ are the equilibrium major radius and poloidal magnetic field,
respectively.

The above factorization contains three applications of $\bb{E}^{-1}$; however, it was
found that only the second had a significant impact on the convergence rate, and so the
first and last were omitted, significantly speeding the calculation. The resulting
preconditioning operator can be divided into three stages, $\bb{P} = \bb{P}_3 \bb{P}_2
\bb{P}_1$, and written as an operation on a unpreconditioned state vector $v^{(U)}$, which
produces a state vector $v^{(P)} = \bb{P} v^{(U)}$. In the following description,
superscripts on state vectors or components of state vectors will indicate the stage of
the preconditioning operation.

Note that in the simplest backwards Euler implicit scheme when $\bb{P} = \bb{A}^{-1}$, the
vector $v^{(U)}$ is the solution at the last timestep, and $v^{(P)}$ the solution at the
next.  The preconditioner can therefore be interpreted as a predictor step, calculating an
approximation of the state vector at the next timestep from the previous one. The
application of $\bb{P}$, Equation \eqref{eq:factorisation}, is performed as follows.
\begin{enumerate}
  \item Given an unpreconditioned vector $v^{(U)} = \left(\rho^{(U)}, T_i^{(U)},T_e^{(U)},
    v_{||}^{(U)}, A_{||}^{(U)}, U^{(U)}\right)$, apply the first matrix in
    Equation~\eqref{eq:factorisation}:
    \[
      v^{(P1)} = \boutmatrix
                  {\bb{I}} {0}
                  {-\bb{L}} {\bb{I}}
                v^{(U)} ,
    \]
    where the $\bb{E}^{-1}$ operator has been dropped, as discussed above. When this is
    expanded, only the vorticity field is modified:
    \[
      U^{(P1)} = U^{(U)} + \gamma \frac{B_0}{\rho_0}\nabla_{||}J_{||}^{(U)}
              + \gamma \frac{1}{\rho_0}2\vec{b}\times\kappa\cdot\nabla P^{(U)} ,
    \]
    where $J_{||}^{(U)} = -\nabla_\perp^2A_{||}^{(U)}$ and $P^{(U)} =
    \rho_0\left(T_i^{(U)} + T_e^{(U)}\right) + \rho^{(U)}\left(T_{i0} +
      T_{e0}\right)$. This operation represents a forward Euler step to predict the
    vorticity after a timestep $\gamma$.

  \item Apply the second matrix, solving for parallel temperature conduction and shear
    Alfv\'en wave dynamics,
    \[
    v^{(P2)} = \boutmatrix
                {\bb{E}^{-1}} {0}
                {0} {\bb{P}_{Schur}^{-1}}
              v^{(P1)} ,
    \]
    which can be written as
    \begin{align*}
      T_{s}^{(P2)} &= \left(
                       \bb{I} - \gamma \frac{\chi_{||s}}{\rho_0}\nabla^2_{||}
                     \right)^{-1}T_{s}^{(U)} \cr
      U^{(P)} &= \left(
                  \bb{I} + \frac{2\gamma^2\nabla_{||}(RB_\theta)}{RB_\theta B_0^2}
                  - \gamma^2\frac{B_0^2}{\rho_0}\nabla_{||}^2
                \right)^{-1}U^{(P1)} .
    \end{align*}
    Since all these operators are of the form $\left(a + b\nabla_{||}^2\right)^{-1}$ with
    different scalars $a$ and $b$, the same parallel solver can be used. Since the
    equations for temperature and vorticity are decoupled, a further optimization, not
    used in this current work, would be to also solve the three equations at the same
    time.
    This operation represents a backward Euler step for the propagation of vorticity and
    temperature along magnetic fieldlines over a timestep $\gamma$ and is the most
    computationally demanding.  Efficiently solving these systems is critical to the
    performance of the preconditioner, so the method used is described in detail below.

  \item Apply the third matrix, updating the other fields based on the new vorticity
    \[
      v^{(P)} = \boutmatrix
                 {\bb{I}} {-\bb{U}}
                 {0} {\bb{I}}
               v^{(P2)} ,
    \]
    where again the $\bb{E}^{-1}$ operator has been dropped. This can be written as
    \begin{align*}
      \rho^{(P)} &= \rho^{(U)} - \gamma\vec{v}_{E\times B}^{(P)}\cdot\nabla\rho_0 \cr
      T_s^{(P)} &= T_s^{(P2)} - \gamma \vec{v}_{E\times B}^{(P)}\cdot\nabla T_{s0} \cr
      A_{||}^{(P)} &= A_{||}^{(U)} - \gamma\nabla_{||}\phi^{(P)} ,
    \end{align*}

    where $\vec{v}_{E\times B}^{(P)} = \frac{1}{B_0} \vec{b}_0\times\nabla\phi^{(P)}$ is
    the perpendicular $E\times B$ velocity calculated by using $\phi^{(P)} =
    \nabla_\perp^{-2}\left(B_0 U^{(P)}\right)$, the electrostatic potential calculated
    from the updated vorticity.  This represents an implicit update of $\rho,T_s,A_{||}$,
    advecting plasma quantities based on an approximate $E\times B$ velocity at the next
    timestep.

    The only variable that is not modified in any of these operations is the parallel
    velocity, so $v_{||}^{(P)} = v_{||}^{(U)}$.
\end{enumerate}

The preconditioner described above contains the key physics responsible for fast
timescales in the system, and it has greatly reduced the complexity and computational cost
of calculating $\bb{P}$.  These features can be seen by considering the sizes of the
matrices involved.  Since 6 variables are evolved on $N$ mesh points, the matrix $\bb{A}$
to be inverted has $(6N)^2$ elements, most of which are however zero. A typical simulation
will use on the order of 100 points in each dimension (between 64 and 512 in most cases,
though the radial mesh size $N_\psi$ can be several thousand for ITER simulations) and so
have $\mysim 10^6$ mesh points, requiring the inversion of a $\left(6\times 10^6\right)^2$
matrix at each timestep. By block factorizing and keeping only the fast components, we
have reduced this to solving three separate matrices (in $\bb{E}$ and $\bb{P}_{Schur}$),
which can be performed in parallel, each of which is of size $(N)^2 \simeq (10^6)^2$. This
approach in itself does not represent a large savings; but by exploiting the structure of
these blocks in the field-aligned coordinate system employed by BOUT++, further
simplifications can be made.

Since each block contains only derivatives along magnetic fieldlines after we remove
perpendicular derivatives, and equilibrium magnetic fieldlines form helices around a torus
covering 2D ``flux'' surfaces, we can independently solve for each flux surface,
immediately reducing the $(10^6)^2$ inversion problem to $\mysim 100$ separate $(10^4)^2$
problems. This assumption of equilibrium flux surfaces will break down if the magnetic
field becomes significantly perturbed in the later stages of an ELM crash, but this is not
a critical problem because at these times nonlinear dynamics are fast and a smaller
timestep is unavoidable.

To further simplify the problem, we exploit the toroidal symmetry of the equilibrium and
decompose into toroidal Fourier harmonics. This approach decouples the toroidal and
poloidal directions, splitting the 2D problem into a set of one-dimensional equations in
poloidal angle, with a complex phase shift representing the pitch of the fieldlines. If we
started with $N_\psi$ flux surfaces, each containing $N_\theta$ poloidal points and
$N_\zeta$ toroidal points (so $N = N_\psi\times N_\theta\times N_\zeta$), then we are left
with $N_\psi\times \left(N_\zeta/2\right) \sim 10^4$ complex cyclic tridiagonal systems,
each of length $N_\theta \sim 10^2$. These are then solved efficiently in parallel by
using a variant of the cyclic reduction algorithm with interface equations
\cite{austin2004}. Since the three equations being solved are also independent, all
$3\times N_\psi\times \left(N_\zeta/2\right)$ tridiagonal systems can be solved
simultaneously, resulting in a good overlap of communication and computation on a large
number of processors.

\subsection{Nonlinear GMRES}
\label{subsec:ngmres}

As an alternative to preconditioned Newton-Krylov methods, we consider nonlinear GMRES
(NGMRES) to solve the nonlinear system~\eqref{eq:f=0}.  NGMRES and other similar methods
(quasi-Newton, Anderson mixing) of solution for nonlinear systems are analogous to
multistep Krylov methods used for linear systems \cite{Oosterlee:2000tg,Anderson:1965wp}.
At each iteration, an updated solution is constructed as a linear combination of several
stored previous solutions and a new candidate solution.  A minimum-residual linear
combination is found by solving a small minimization problem.

The new candidate solution is constructed by perturbing the current solution by some
negative damping of the current residual, typically found through a line search.  One may
also construct the candidate using a small number of iterations of some other nonlinear
solution technique. For the present problem we use a damping factor of $-0.001$ on the
residual to construct the candidate and store 50 previous solutions. By using these Krylov
subspace accelerators, we aim to reduce the overall CPU time for each nonlinear solve and
also reduce memory usage.

\section{Experimental Results}
\label{sec:results}

We use the nonlinear solvers in the PETSc library \cite{petsc-user-ref,efficient} to solve
these nonlinear systems over varying numbers of processors.  The library provides a simple
interface for nonlinear solvers that enables selection of algorithms and parameters at
runtime, with no changes to the application code.

Our experiments were run on the Fusion cluster at Argonne National Laboratory.  This
cluster has 320 compute nodes, each with an Intel Nehalem dual-socket, quad-core 2.53 GHz
processor (or 2,560 compute processors), connected by Infiniband.

We specified a relative nonlinear convergence tolerance of $10^{-3}$ for all methods,
which is consistent to what is typically employed for implicit timestepping.  For each
Newton-Krylov run, we used GMRES with a restart of 30 and a relative linear convergence
tolerance of $10^{-2}$.  The method NGMRES, by definition, has no linear solver associated
with it.  Consequently, each nonlinear solve using NGMRES has many more iterations, in
contrast with the Newton-Krylov approaches, where an average of two or three nonlinear
iterations achieved convergence.

\subsection{Convergence of Nonlinear Solvers}
\label{subsec:convergencenonlinear}

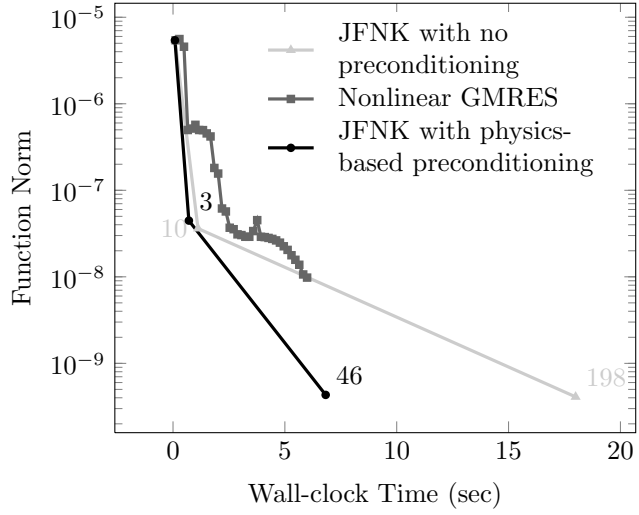
\begin{figure}[ht]
  \centering
  \begin{tikzpicture}
    \begin{semilogyaxis}[
      ylabel={Function Norm},
      xlabel={Wall-clock Time (sec)},
      cycle list name=bwcolors,
      legend style={
        legend columns=1,      % override the global setting
        align=left,            % warning: this option enables multiple lines in the legend
        nodes=right,
      },]
      \addplot table {reconnect-2field-none-16.snes};
      \addplot table {reconnect-2field-ngmres-16.snes};
      \addplot table {reconnect-2field-precon-16.snes};

      % without an 'align={left, right, or center}' then the following will break because
      % of '\\'
      \addlegendentry{JFNK with no\\ preconditioning};
      \addlegendentry{Nonlinear GMRES};
      \addlegendentry{JFNK with physics-\\based preconditioning};
    \end{semilogyaxis}
  \end{tikzpicture}
  \caption{Function norm ($l_2$) versus wall-clock time (sec) on 16 cores for two-field
    test case, Equation~\ref{eq:2field}. Labels on the figure indicate cumulative number
    of linear solver iterations for Newton-Krylov methods.}
  \label{fig:intro-convergence-16}
\end{figure}

To test the feasibility of our approach to physics-based preconditioning, we conducted
initial experiments with a simpler problem: a two-field subset of the six-field model that
contains only the Alfv\'en wave,
\begin{align}
  \pd{A_{||}}{t} &= -\nabla_{||}\phi - \eta J_{||} \cr
  \pd{U}{t} &= -\vec{v}_{E\times B}\cdot\nabla U
             +\frac{1}{\rho_0} B_0^2\nabla_{||} \left(\frac{J_{||} + J_{||0}}{B_0}\right) ,
  \label{eq:2field}
\end{align}
with auxiliary equations
\[
  U = \frac{1}{B}\nabla_\perp^2\phi \qquad J_{||} = -\nabla_\perp^2 A_{||} .
\]
Though greatly simplified, this model can describe the essential physics of magnetic
reconnection and island formation in tokamak plasmas~\cite{fitzpatrick1998}. Thus, the
efficient solution of this system of equations is of interest in itself. We used a
rectangular (slab) grid of dimension $68\times 32 \times 16$.  As seen in
Figure~\ref{fig:intro-convergence-16}, physics-based preconditioned JFNK showed a significant
reduction in overall runtime in comparison with the default (JFNK with no
preconditioning). NGMRES also significantly decreased overall runtime.

For the remaining experiments, we used the six-field model, Equations
\eqref{eq:6field1}-\eqref{eq:6field3}, on a grid size of $N_\psi=64$, $N_\theta=64$, and
$N_\zeta=128$, partitioned only in the $\psi\theta$-plane because of using the discrete
Fourier transform in the $\zeta-$direction. Each example has six coupled independent
variables $(\rho, T_e, T_i, U, v_{||}, A_{||})$, as given by Equation~\eqref{eq:f=0}. One
reason for using a relatively small grid is to test the scaling of these algorithms when
the number of grid cells per processor becomes small. Since domain decomposition is in
$\psi$ and $\theta$, 512 processors (the maximum used here) will divide the mesh into $4\times 2\times 128$
blocks. This is a more useful test than simply using a large mesh to demonstrate scaling.

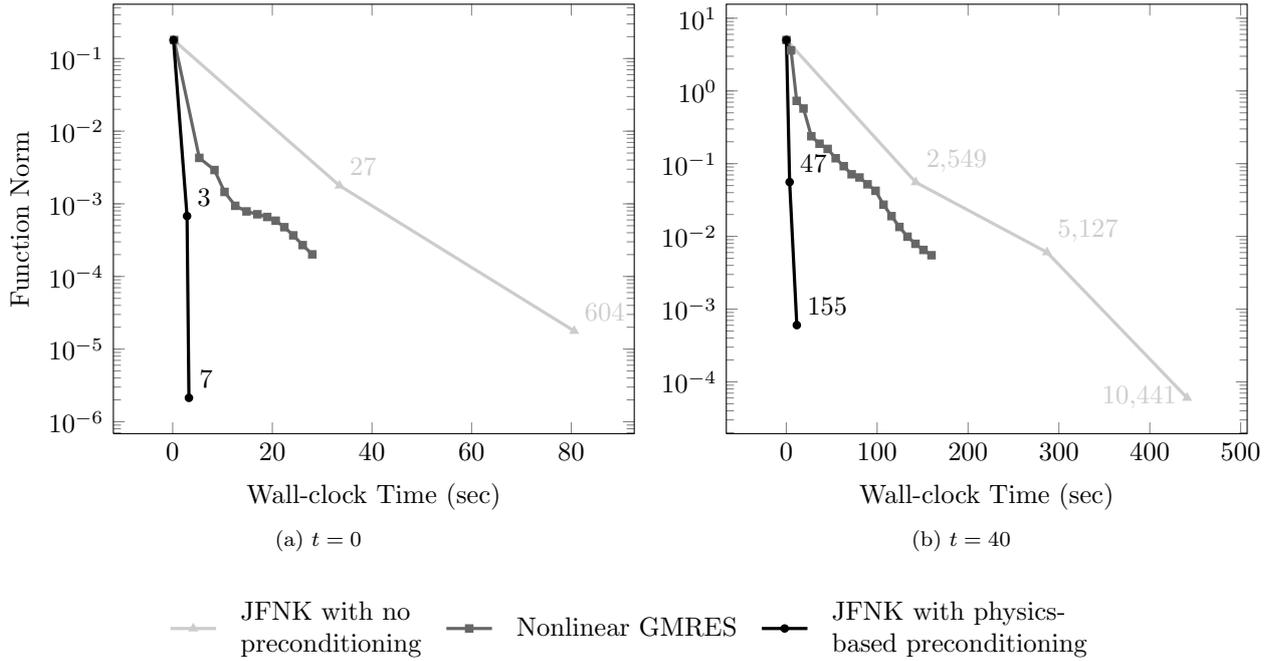
\begin{figure}[ht!]
  \centering
  \subfloat[$t=0$]{
    \begin{tikzpicture}
      \begin{semilogyaxis}[
        ylabel={Function Norm}, % only adding a y-label to the first plot since they are
                                % side-by-side
        xlabel={Wall-clock Time (sec)},
        legend to name=convergence-legend,
        cycle list name=bwcolors,
        ]
        \addplot table {t0-none-512.snes};
        \addplot table[each nth point={10}] {t0-ngmres-512.snes};
        \addplot table {t0-precon-512.snes};

        % without an 'align={left, right, or center}' then the following will break
        % because of '\\'
        \addlegendentry{JFNK with no\\ preconditioning};
        \addlegendentry{Nonlinear GMRES};
        \addlegendentry{JFNK with physics-\\based preconditioning};
      \end{semilogyaxis}
    \end{tikzpicture}
    \label{fig:convergence0}
  }
  \subfloat[$t=40$]{
    \begin{tikzpicture}
      \begin{semilogyaxis}[
        xlabel={Wall-clock Time (sec)},
        cycle list name=bwcolors,
        ]
        \addplot table {t40-none-512.snes};
        \addplot table[each nth point={40}] {t40-ngmres-512.snes};
        \addplot table {t40-precon-512.snes};
      \end{semilogyaxis}
    \end{tikzpicture}
    \label{fig:convergence40}
  }\\
  \vspace{3ex}
  \hspace{3ex}
  \pgfplotslegendfromname{convergence-legend}
  \caption{Function norm ($l_2$) versus wall-clock time (sec) on 512 cores for the full
    six-field model (Section~\ref{subsec:equations}) at fixed problem size.  {\em Left:}
    Results for the initial timestep $t=0$. {\em Right:} Results for timestep $t=40$.
    Labels indicating the cumulative number of linear solver iterations for Newton-Krylov
    methods show that the linearized Newton system is relatively easy to solve initially,
    though it becomes more difficult at timestep $t=40$.}
\end{figure}

As shown in Figures \ref{fig:convergence0} and \ref{fig:convergence40}, physics-based
preconditioning achieves an order of magnitude decrease in overall time for the
Newton-Krylov nonlinear solve and nearly two orders of magnitude reduction in the number
of Krylov iterations in comparison with the unpreconditioned variant.  NGMRES also
significantly reduces overall time for the nonlinear solve compared with the BOUT++
default of JFNK with no preconditioning.  Note that while the nonlinear relative
convergence tolerance for each run is $10^{-3}$, the final value of the function norm for
the Newton-Krylov runs is much smaller compared with the NGMRES run because of the
substantial decrease in function norm at each Newton step.  While the physics-based
preconditioner code is hand-written for each type of physics model, NGMRES requires no
application-specific code and can be activated at runtime, making it an attractive
alternative for efficient and scalable nonlinear solves.

At early simulation times ($t=0$, Figure~\ref{fig:convergence0}), nonlinear terms in the
evolution equations are negligible, and so the solution is essentially exponential
growth. At later times ($t=40 / \omega_A$, Figure~\ref{fig:convergence40}), nonlinear
terms become important, filaments begin to erupt outwards from the plasma, and the
solution becomes more complicated, even turbulent.  As a result the number of iterations
and wall clock time increase significantly for later timesteps.  The physics-based
preconditioner is marginally less effective at later times, reducing the iteration count
over unpreconditioned JFNK by a factor of $67$ at $t=40$, compared with a factor of $86$
at $t=0$. This result is expected because nonlinear terms were not included in
construction of the preconditioner.  Nevertheless, the linear preconditioner remains
highly effective, indicating that linear dynamics are still limiting the timestep even
when the plasma evolution is nonlinear.

\subsection{Scalability of Nonlinear Solvers}
\label{subsec:scalabilitynonlinear}

% ran the following command (one line; just copy and paste in the same directory with all
% the *.snes files)
%
% echo -e "procs, 16, 32, 64, 128, 256, 512," "$(for p in {none,ngmres,precon}; do echo -ne "\n$p, "; for i in {16,32,64,128,256,512}; do cat $p-$i.snes | awk -F', ' 'END {print $3", "}' | xargs echo -n; echo -n ' '; done; done)" | awk -F', ' '{for (f = 1; f <= NF; f++) a[NR, f] = $f} NF > nf { nf = NF } END {for (f = 1; f <= nf; f++) for (r = 1; r <= NR; r++) printf a[r, f] (r==NR ? RS : FS)}'

% store the output of the above command in the \scalability macro, remember the comma
% separator was set above
\pgfplotstableread{
procs, none, ngmres, precon
16, 6.079748e+03, 1.349796e+03, 1.431308e+02
32, 3.624823e+03, 8.953517e+02, 9.952766e+01
64, 1.685646e+03, 4.164732e+02, 4.088985e+01
128, 9.180763e+02, 2.658314e+02, 2.383816e+01
256, 5.977239e+02, 1.971505e+02, 1.552403e+01
512, 4.410280e+02, 1.653697e+02, 1.161303e+01
}\scalability

\begin{figure}[ht!]
  \centering
  \begin{tikzpicture}
    \begin{groupplot}[
      group style=
      {
        group size=2 by 2,
        xlabels at=edge bottom,
        ylabels at=edge left,
        horizontal sep=0.075\textwidth,
        group name=scalability
      },
      scalabilitybar/.style={
        ybar,
        ymin=0,
        cycle list name=bwcolors-bar,
      },
      xtick=data,
      enlarge x limits=0.2,
      xticklabels from table={\scalability}{procs},
      xmode=log,
      ylabel={Wall-clock Time (sec)},
      xlabel={Number of Processors},
      table/x=procs,
      legend style={
        draw=none,
        nodes=right,
        align=left,
        /tikz/every even column/.append style={column sep=5pt},
      },
      ]

      \nextgroupplot[scalabilitybar]
      \pgfplotsset{cycle list shift=3}
      \addplot table[y=none] {\scalability};
      \pgfplotsset{cycle list shift=-3}
      \addplot table[y=ngmres] {\scalability};
      \pgfplotsset{cycle list shift=-2}
      \addplot table[y=precon] {\scalability};
      \addlegendentry {JFNK with no\\ preconditioning};
      \addlegendentry {Nonlinear GMRES};
      \addlegendentry {JFNK with physics-\\based preconditioning};

      \nextgroupplot[scalabilitybar]
      \pgfplotsset{cycle list shift=2}
      \addplot table[y=ngmres] {\scalability};
      \addlegendentry {Nonlinear GMRES};

      \nextgroupplot[scalabilitybar]
      \addplot table[y=precon] {\scalability};
      \addlegendentry {JFNK with physics-\\based preconditioning};

      \nextgroupplot[
        ymode=log,
        log base 10 number format code/.code={
          $\pgfmathparse{10^(#1)}\pgfmathprintnumber{\pgfmathresult}$
        },
        cycle list name=bwcolors,
        legend pos=south west,
        legend style={
          nodes=right,
          align=left,
          fill=none,
          font=\footnotesize,
        }
      ]

      \addscalingplot [table/y=none] {\scalability};
      \addscalingplot [table/y=ngmres] {\scalability};
      \addscalingplot [table/y=precon] {\scalability};
      \addlegendentry {JFNK with no\\ preconditioning};
      \addlegendentry {Nonlinear GMRES};
      \addlegendentry {JFNK with physics-\\based preconditioning};
    \end{groupplot}
  \end{tikzpicture}
  \caption{The upper left plot compares the scalability of each algorithm for the full
    six-field model.  The upper right and lower left plots are close-ups of NGMRES and
    physics-based preconditioning, respectively. The lower right plot compares each
    algorithm with ideal scaling. We can see that NGMRES and especially physics-based
    preconditioning significantly reduced overall runtime.}
  \label{fig:scalability}
\end{figure}
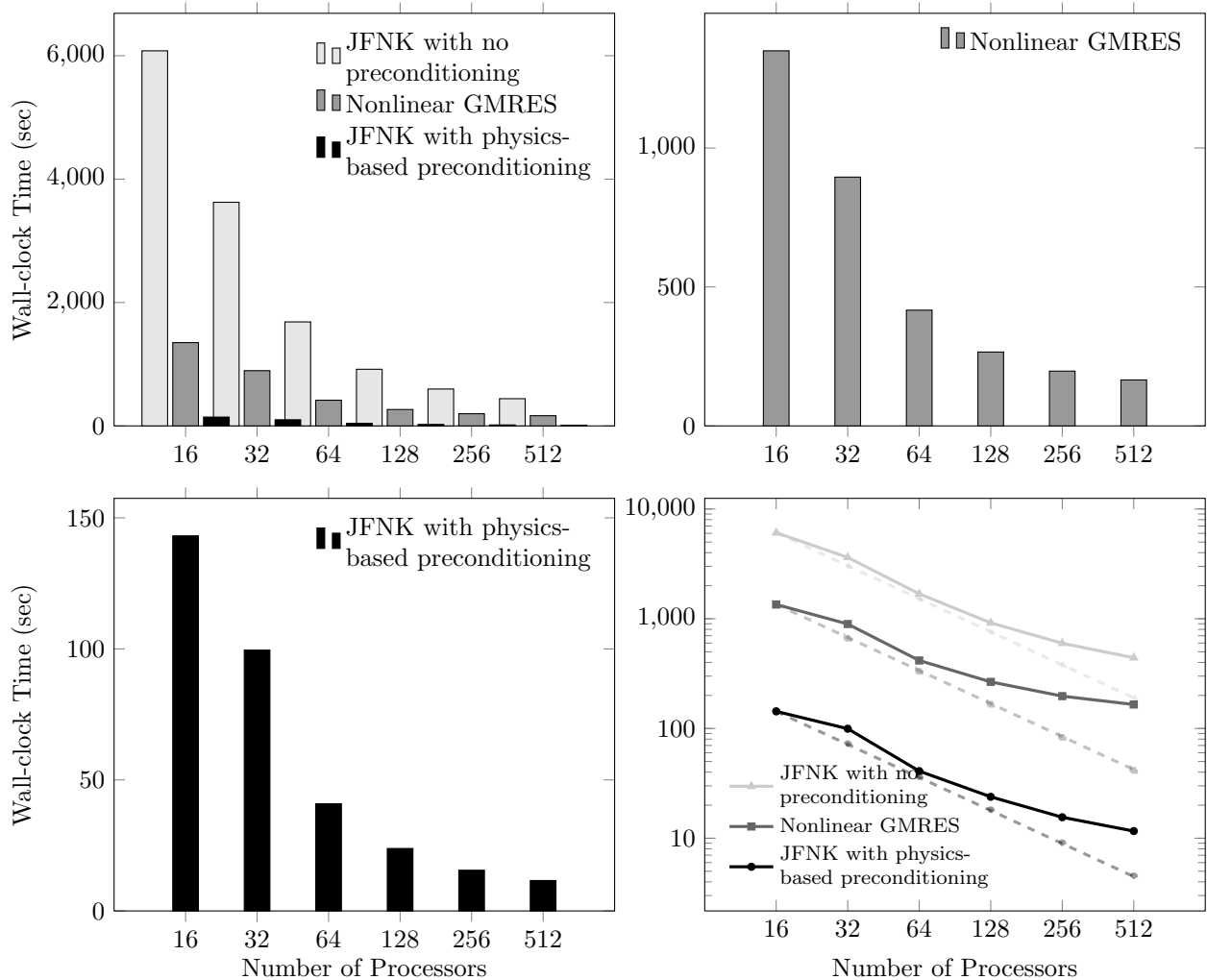

\pgfplotstableread{
procs, rhs, VecMDot, VecMAXPY
16, 5.9893e+03, 1.0253e+03, 3.9946e+01
32, 3.5834e+03, 5.1720e+02, 1.7514e+01
64, 1.6591e+03, 2.3304e+02, 9.9039e+00
128,9.0413e+02, 1.3117e+02, 4.8912e+00
256,5.8619e+02, 8.6498e+01, 2.9258e+00
512,4.3136e+02, 8.1301e+01, 1.5468e+00
}\logsummarytimenone

\pgfplotstableread{
procs, rhs, VecNorm, VecDot, VecMDot
16, 1.1648e+03, 1.9524e+02, 3.8848e+01, 2.3147e+01
32, 7.7791e+02, 1.4300e+02, 2.8514e+01, 1.0978e+01
64, 3.2883e+02, 7.6011e+01, 2.6913e+01, 5.9854e+00
128, 1.8896e+02, 5.6367e+01, 2.7379e+01, 3.4278e+00
256, 1.1926e+02, 5.0114e+01, 3.0705e+01, 2.1295e+00
512, 8.4222e+01, 5.1892e+01, 3.4524e+01, 1.3097e+00
}\logsummarytimengmres

\pgfplotstableread{
procs, rhs, PCApply, VecMDot, VecNorm
16, 1.2007e+02, 2.1316e+01, 1.6914e+01, 1.3464e+00
32, 7.9991e+01, 1.8606e+01, 1.1552e+01, 9.6238e-01
64, 3.3514e+01, 6.7648e+00, 4.7187e+00, 4.3807e-01
128,1.9236e+01, 4.0946e+00, 2.8131e+00, 3.1593e-01
256,1.2155e+01, 2.8135e+00, 1.9681e+00, 2.8729e-01
512,8.5291e+00, 2.4913e+00, 1.7678e+00, 3.2046e-01
}\logsummarytimeprecon

\begin{figure}[ht!]
  \centering
  \includegraphics{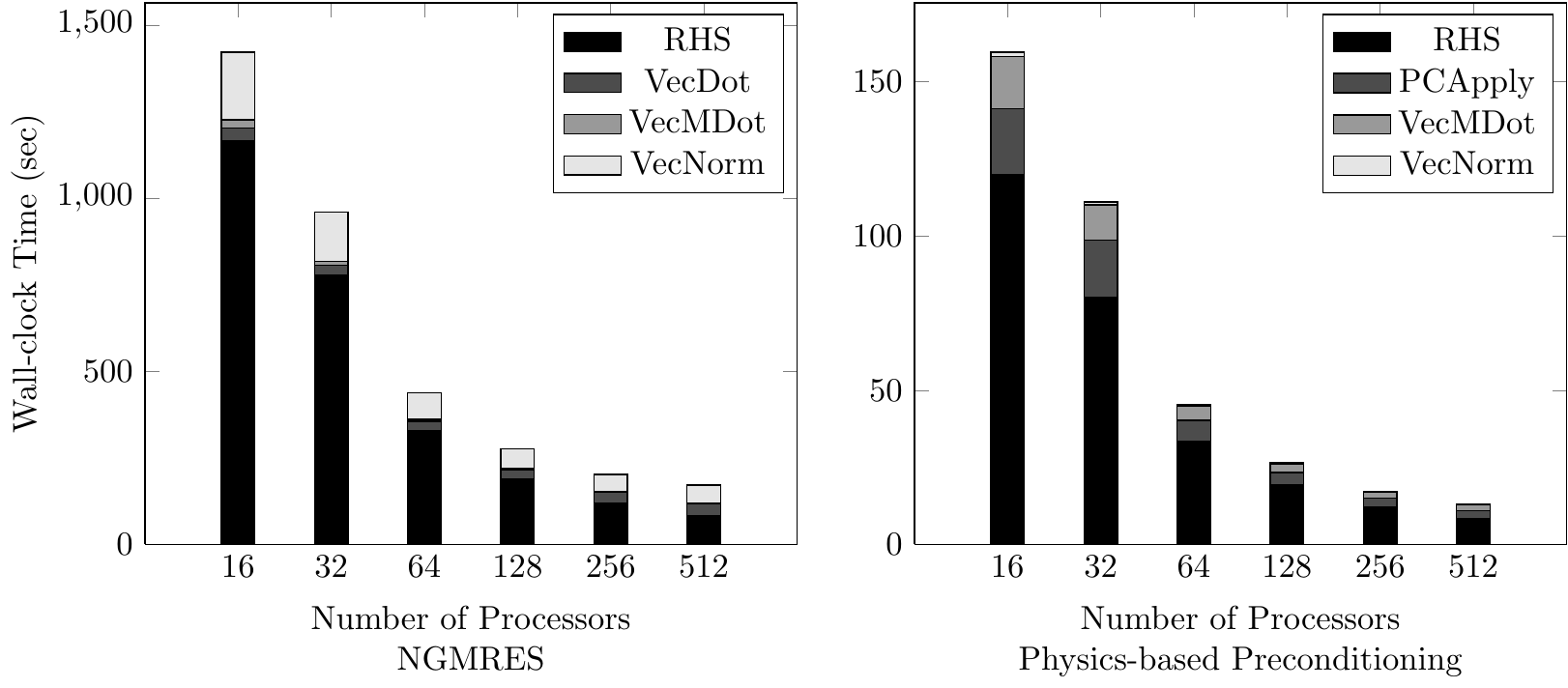}
  \caption{Results showing that function evaluation (RHS) dominates the time for the
    six-field model}
  \label{fig:stacked}
\end{figure}

Figure~\ref{fig:scalability} shows the scaling of the JOREK test case with the Jacobian-free
Newton-Krylov (no preconditioning), nonlinear GMRES, and Jacobian-free Newton-Krylov (with
physics-based preconditioning) algorithms in the first subfigure with lighter bars
denoting the comparison with NGMRES and physics-based preconditioning.  Notice that the
physics-based preconditioning is barely legible in the first subfigure because of the
overall runtime being an order of magnitude faster than the default solver.  For this
reason the individual plots of NGMRES and physics-based preconditioning are also displayed
as individual subfigures.  All three algorithms then are plotted in a log-log graph in
conjunction with ideal scaling (the dashed lines). One can see that all three algorithms
have almost identical scaling.  This relatively small test case becomes too small for core
counts larger than 128, as evidenced by the tapering of scalability.

Clearly, using a physics-based preconditioner achieves the fastest runtime.  Physics-based
approaches, however, require detailed knowledge of the modeling equations being used
\cite{chacon2008,chacon02,ChaconKnollEtAl02}, enforcing a tight coupling between the
programmer and physics-based preconditioning analysis. Even if such a coupling exists,
potential challenges arise with implementation because of complexities of physics and
preconditioning algorithms.  In contrast, NGMRES requires no application-specific
customization, since the application provides code only for the nonlinear function
evaluation.

The bar graphs in Figure~\ref{fig:stacked} show the proportion of time spent in various
phases of the nonlinear solve.  For each run, the evaluation of the function (or right-hand
side, RHS) dominates the runtime. For Jacobian-free variants, a function evaluation is
required for each Jacobian-vector product (one per Krylov iteration).  The preconditioned
variant of JFNK requires fewer function evaluations because of better-conditioned linear
systems.  NGMRES, alternatively, needs roughly ten times more function evaluations than
does preconditioned JFNK. Also, we can see that the time needed for computing the search
direction in NGMRES (VecNorm) is nontrivial. This result further indicates that the
parallel scaling shown in Figure~\ref{fig:scalability} is due primarily to the scaling of the
nonlinear function evaluation; improving this scaling is the subject of ongoing work.

\section{Conclusions}
\label{sec:conclusions}

We have introduced a physics-based preconditioner that leverages insight into the physics
of the equations being solved in order to improve convergence of matrix-free Newton-Krylov
methods in BOUT++ ELM simulations.  We also introduced the use of a complementary
approach, nonlinear GMRES, which requires no Jacobian-vector products or preconditioning.
We demonstrated that both approaches reduce time for solving the nonlinear systems that
arise at each timestep of an implicit ELM simulation.

The physics-based preconditioning scheme described in this paper is straightforward to
construct once the analytic Jacobian is known, can be efficiently solved in magnetic
field-aligned coordinate systems commonly used in plasma simulations, and is highly
effective for both hyperbolic (Alfv\'en wave) and parabolic (heat conduction) problems.
The preconditioning steps have intuitive interpretations, making it straightforward to
experiment with adding and removing terms in the equations in order to optimize the
preconditioner.  By applying this method to a six-field reduced MHD model,
order-of-magnitude reductions in wall clock time have been achieved, even in regimes where
nonlinear terms in the evolution equations are important.  The application of this
improved code to ELM simulations, and predictions for ITER, will be reported in a separate
paper.

Using a relatively small problem size, we have demonstrated good scaling up to 512
processors. Since the costly matrix inversion step in the preconditioner is trivially
parallelized in two dimensions ($\psi$ and $\zeta$), increasing the problem size should
allow scaling to a much larger number of processors. Testing this will be the subject of
future work.

Other directions of future work include the development of additional physics-based
preconditioning variants, such as fieldsplits for fast Alfv\'en waves and thermal
conductivity along the fieldlines.  We are also exploring broader issues in
time-dependent simulations, including the use of implicit-explicit (IMEX) approaches for
flexible timestepping. Work is under way to incorporate BOUT++ into the FACETS
framework~\cite{facets-scidac09,facets-fec2010} in order to facilitate research on
core-edge-wall coupling.

\section*{Acknowledgments}

We gratefully acknowledge use of the Fusion cluster in the Laboratory Computing Resource
Center at Argonne National Laboratory.  S. Farley and L. McInnes were supported by the
Office of Advanced Scientific Computing Research, Office of Science, U.S. Department of
Energy, under Contract DE-AC02-06CH11357. B. Dudson was supported by EFDA WP11 and HPC-FF
computing resources. This work was carried out largely as part of the FACETS SciDAC
project supported by DOE. We also thank L.Chacon for his presentation and insightful
discussions at the 2011 BOUT++ workshop, which provided the impetus for this work.

\bibliographystyle{elsarticle-num}
\bibliography{paper}

%\newpage
\vspace{2cm}
{\bf Government License.} The submitted manuscript has been created by UChicago
Argonne, LLC, Operator of Argonne National Laboratory (``Argonne").  Argonne, a
U.S. Department of Energy Office of Science laboratory, is operated under Contract
No. DE-AC02-06CH11357. The U.S. Government retains for itself, and others acting on its
behalf, a paid-up nonexclusive, irrevocable worldwide license in said article to
reproduce, prepare derivative works, distribute copies to the public, and perform publicly
and display publicly, by or on behalf of the Government.

\end{document}